\documentclass[12pt,preprint]{aastex}
\usepackage{colordvi}
\usepackage{color}
\begin{document}

\title{COMPARISON OF EMISSION PROPERTIES OF TWO HOMOLOGOUS FLARES IN AR 11283}

\author{Yan Xu,
        Ju Jing,
        Shuo Wang,
        and Haimin Wang}

\affil{ Space Weather Research Lab, Center for Solar-Terrestrial
Research, \\ New Jersey Institute of Technology\\
323 Martin Luther King Blvd, Newark, NJ 07102-1982}

\email{yx2@njit.edu}


\begin{abstract}

Large, complex, active regions may produce multiple flares within a certain period of one or two days. These flares
could occur in the same location with similar morphologies, commonly referred to as ``homologous flares''. In 2011
September, active region NOAA 11283 produced a pair of homologous flares on the 6th and 7th, respectively. Both
of them were white-light (WL) flares, as captured by the Helioseismic and Magnetic Imager (HMI) onboard the
Solar Dynamics Observatory in visible continuum at 6173~\AA\ which is believed to originate from the deep solar
atmosphere.We investigate the WL emission of these X-class flares with HMI’s seeing-free imaging spectroscopy.
The durations of impulsive peaks in the continuum are about 4 minutes. We compare the WL with hard X-ray
(HXR) observations for the September 6 flare and find a good correlation between the continuum and HXR both
spatially and temporally. In absence of RHESSI data during the second flare on September 7, the derivative of
the GOES soft X-ray is used and also found to be well correlated temporally with the continuum. We measure
the contrast enhancements, characteristic sizes, and HXR fluxes of the twin flares, which are similar for both
flares, indicating analogous triggering and heating processes. However, the September 7 flare was associated with
conspicuous sunquake signals whereas no seismic wave was detected during the flare on September 6. Therefore,
this comparison suggests that the particle bombardment may not play a dominant role in producing the sunquake
events studied in this paper.

\end{abstract}


\section{Introduction}

Observations and modeling have demonstrated that flare energy is released in current sheets where magnetic reconnection occurs (e.g., recent review by \citealt{Hudson2011}). As a consequence, particles, including electrons and ions, can be accelerated and propagate upward along the open field lines or spiral downward along the closed field lines. The latter group of energetic particles can penetrate down to the chromosphere or even photosphere and generate flare footpoint emissions in HXR and visible continua, which is also known as white-light (WL). It is hard to detect the WL signal during a flare because the flare emission is much weaker than the background of solar radiation. Apparently, most WL emissions are identified in large flares, although they are believed to exist in all flares \citep{Neidig1989, Zirin1988}. During the era dominated by ground-based observations, only about 120 WL flares were reported and most of them were above X2 class \citep{Hudson2006, Neidig1993b}. \citet{Wang2009} systematically investigated all WLFs with Hinode observations in the G-band at 4305~\AA, and found that the cut-off visibility was reached for M1 flares. Using the 1-meter ground-based telescope, \citet{Jess2008} detected the WL emission from a C2 flare. There are limitations on observing WL flares, such as the observing durations, spatial/temporal resolutions, limited field-of-view (FOV), dynamic range of detectors, choice of filters (wavelengths) and seeing conditions for ground-based observations. The newly launched space telescope, Solar Dynamics Observatory (SDO) \citep{SDO}, provides full disk and imaging spectroscopy capabilities and hence increases the chance of catching WL flares.

A substantial amount of work has been undertaken in an attempt to understand the energy transport and release processes. Two fundamental questions need to be answered: 1) Where does the WL emission originate? and 2) What is the energy source of the WL emission? In the literature, two groups of models have been proposed to address these issues concerning WL flares. Considering the fact that WL emission is always associated with HXR emission, the direct heating model, a straight forward model, assumes that the accelerated electrons reach the deeper atmosphere and deposit their energy by collision \citep{Najita1970, Hudson1972, Ding2003b}. According to \citet{Vernazza1981}, the photospheric column density exceeds some $10^{23}$ cm$^{−2}$ implying that only electrons with initial energy higher than 600 keV can contribute to the heating of the photosphere \citep{Xu2012a}. However, there is no sufficient electron flux derived from HXR observations reaching the $\tau_{5000} = 1$ level. Given the energetic difficulties associated with the direct heating model, some models involving secondary effects have been proposed, such as the chromospheric backwarming model \citep{Hudson1972, Aboudarham1986, Metcalf1990} and the H$^{-}$ emission model \citep{Aboudarham1987, Machado1989, Metcalf1990, Metcalf2003, Ding1994, Ding2003b}. Both direct and non-direct heating mechanisms can contribute to a single event. This idea has been demonstrated by the studies of core-halo structures by \citet{Xu2006, Xu2012a} and \citet{Isobe2007}.

A major flare could release energy exceeding 10$^{32}$ erg \citep{Hudson2011}. However, this may be a small amount of the total magnetic free energy stored in a huge and complex Sunspot group. As an evidence, such a strong Sunspot group can produce multiple flares during its life cycle. These flares, which originate in the same site are also known as homologous flares. The observations and analyses of homologous flares can be traced back to the time when the data were recorded in films. \citep{Zirin1967}. \Citet{Zirin1983} presented an observation of a sequence of at least 10 flares occurring in one single active region during a 25-hour window. In the digital era, examples of homologous flares can be found in many studies, \citet[e.g.,][]{Zhang2002, Sui2004, Takasaki2004, Luoni2006, Meshalkina2009, Kumar2010, Chandra2011}. Active region NOAA 10486 produced the famous Halloween events in 2003, among which at least two flares occurred at the same location on 2003 October 29 and November 02 \citep{Xu2006}.

To further understand the energy transport and release processes, it is crucial to investigate the size and brightness of WL flare kernels \citep{Fletcher2007}. Two conjugate ribbons are commonly observed for most flares in H$\alpha$ or EUV wavelengths. In radio and HXR observations, which are the the direct diagnostics of electron beams, footpoint sources are commonly detected except for a few special cases \citep{LiuC2007b} due to relative coarse resolution and non-focus imaging methods. With WL observations, the flare kernels, if detected, are usually fully resolved as footpoint-like cores and ribbon-like halos \citep{Xu2006}. The characteristic size of the flare core could be as small as 0\arcsec.7 in the near Infrared and increases to 2\arcsec\ in G-band. This trend in principle resembles a converging flare loop and favors the direct heating model because other models do not predict such a converging variation \citep{Xu2012a}. Therefore, multi-wavelength observations in WL will provide a diagnosis of direct heating model.

On the other hand, solar flares are not isolated events. They are always associated with other eruptive phenomena, such as filament eruptions, CMEs, Moreton waves and sometimes sunquakes. In fact, all of these phenomena are different manifestations of a single eruptive event \citep{Hudson2011}. Similar to WL flares, sunquakes are photospheric phenomena observed as propagating wavefronts. \citet{Wolff1972} predicted the existence of sunquakes generated by energetic particles during the impulsive phase of flares. \citet{Kosovichev1998} obtained the first observational evidence of sunquakes associated with an X-class flare on 1996 July 9. Further observations were reported by \citet{Besliu-Ionescu2005}; \citet{Donea1999, Donea2005, Donea2006}; \citet{Kosovichev2007} and \citet{Zharkova2007}. Previous observations have shown that the sunquake sources were cospatial with HXR or $\gamma$-ray flare kernels \citep{Besliu-Ionescu2005, Moradi2007, Zharkova2007, Martinez-Oliveros2008}, indicating a close relationship between sunquakes and energetic particles. \citet{Donea2006} found that many sunquakes coincide with WL flares. The authors believed that the back-warming mechanism is not only responsible for the WL flare emission, but also the energy source of sunquakes. Besides the models related to electron beams, some other mechanisms can contribute to generate sunquakes as well. \citet{Zharkova2005, Hudson2008} believe magnetic reconfiguration or perturbation of flux ropes \citep{Zharkov2011} can generate sunquakes.

In this paper, we present the study of a pair of homologous X-class WL flares in 2011 September, observed by the Helioseismic and Magnetic Imager (HMI) \citep{HMI} onboard SDO. These flares occurred close to the disk center and are therefore good candidates for morphological studies. We perform a comprehensive investigation of the first flare (hereafter Flare I) and a comparison to the second flare (hereafter Flare II) associated with a sunquake \citep{Zharkov2013a}, focusing on the following two topics: 1) The basic flare information, such as the contrast enhancement and correlation between WL and HXR emission; 2) The association with sunquakes.

The HMI/SDO observations and data reduction are discussed in \S 2. Detailed analysis is presented in \S 3, followed by the summary and discussion in \S 4.

\section{Observations}

The primary data that we use for this study are images from SDO observations. There are two channels for WL observations available onboard SDO. The broad band observations centered at 4500~\AA\ ($\pm$ 250~\AA) are provided by the Atmospheric Imaging Assembly (AIA) \citep{AIA}. However, this channel is operating in a guiding/alignment mode while providing very low cadence (about one frame per hour), that is not useful for flare studies.

The observations in visible light near 6173~\AA\ presented in this study are based on the second channel, HMI/SDO observations. This instrument is an imaging spectrometer that takes images at six different spectral points at $\pm$ 34~m\AA, $\pm$ 103~m\AA\ and $\pm$ 172~m\AA\ around the Fe I absorption line at 6173.34~\AA\ (referred as 6-point data hereafter). Two types of calibrated data sets are analyzed, namely visible continuum and near real time (NRT) data. {\it The visible continuum images} are actually {\it derived} from the 6-point data by fitting the line-profile of Fe I 6173.34~\AA. During the reconstruction process, one set of 6-point data are integrated using a specific weighting function. Consequently, artificial features may be introduced when generating difference images for flare studies \citep{MartinezOliveros2011}. To avoid ambiguity, we select reference images at least four minutes before the flare and emphasize more on the qualitative analysis of the continuum emission. Nevertheless, the HMI's continuum images are good proxies of WL usually obtained using broad band filters. We will use the words `continuum' and `WL' interchangeably hereafter. The image scale of the WL maps is 0\arcsec.5~per pixel and the effective cadence is about $45$ seconds. The WL data is used for comparison with HXR emission and study of temporal evolution of the flares. {\it The NRT data} contains calibrated 6-point line profiles. Different from WL data, it provides spectral information. HMI has two camera systems, the front one is used to produce the line-of-sight observables by scanning six wavelengths at two polarizations (LCP and RCP). For each scan, twelve images are obtained and are spatially aligned via a linearly interpolation with positive weightings \citep{MartinezOliveros2011, MartinezOliveros2014}. The side camera is used for retrieving the full Stokes vectors. The NRT data investigated in this paper was taken by the front camera after spatial alignment provided by the HMI team. It is used for the multilayer analysis during the flare peak times.

As the direct diagnosis of electron beams, RHESSI \citep{RHESSI} HXR data is used to provide supplementary spatial/temporal information of the footpoints in flare I. RHESSI detects HXR emission using nine rotating modulation collimators (RMC). The rotation period is four seconds, which is basically the shortest time period to obtain an image. The spatial resolution depends on the choice of RMC combinations. By selecting the finest RMC \#1, one can achieve a spatial resolution of 2\arcsec.2 \citep{Dennis2009}. RMCs with larger numbers are thicker and able to absorb more HXR photons to get better statistics for imaging. In this study, RMC 1-7 are selected for HXR imaging using CLEAN method. In absence of RHESSI during the second flare, we use the time derivative of GOES soft X-ray (SXR) light curve as the proxy of the HXR temporal variation.

\section{Results and Analysis}

\subsection{Flare I}

Active region (AR) NOAA 11283 produced two large flares from an identical location on September 6 and 7, respectively. Flare I on September 6 was classified as an X2.1 event. According to GOES SXR record, it was initiated around 22:12 UT and peaked at 22:20 UT. In the HXR, there were two consecutive peaks about three minutes apart. In the continuum, we see two flare kernels which are typical for major flares. At the flare time, the active region was approximately located at N126$\arcsec$W290$\arcsec$ from the disk center.

\begin{figure}
\centering
\includegraphics[scale=1.0]{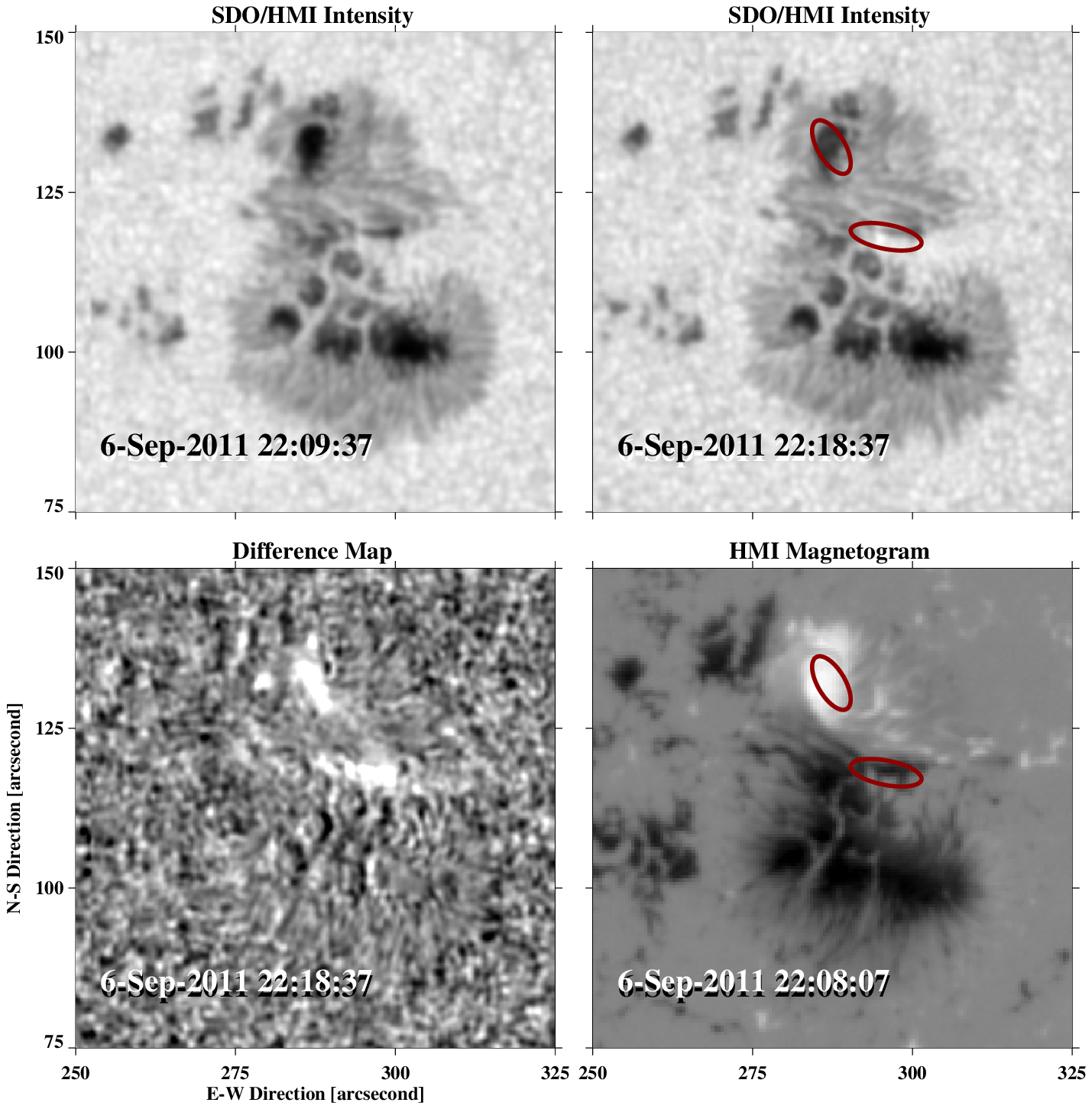}
\caption{WL images and Magnetogram of Flare I on 2011 September 6.
         Upper-left panel: HMI continuum image taken before the flare which is used as the reference frame.
         Upper-right panel: HMI continuum image taken during the X2.1 flare.
         Lower-left panel: Difference image by subtracting the reference frame from the middle panel.
         Lower-right panel: HMI line-of-sight magnetogram taken before the flare at 22:08 UT. The red contours in two right panels show the positions of flare sources relative to the sunspot group. \label{f1}}
\end{figure}

\begin{figure}
\centering
\includegraphics[scale=1.0]{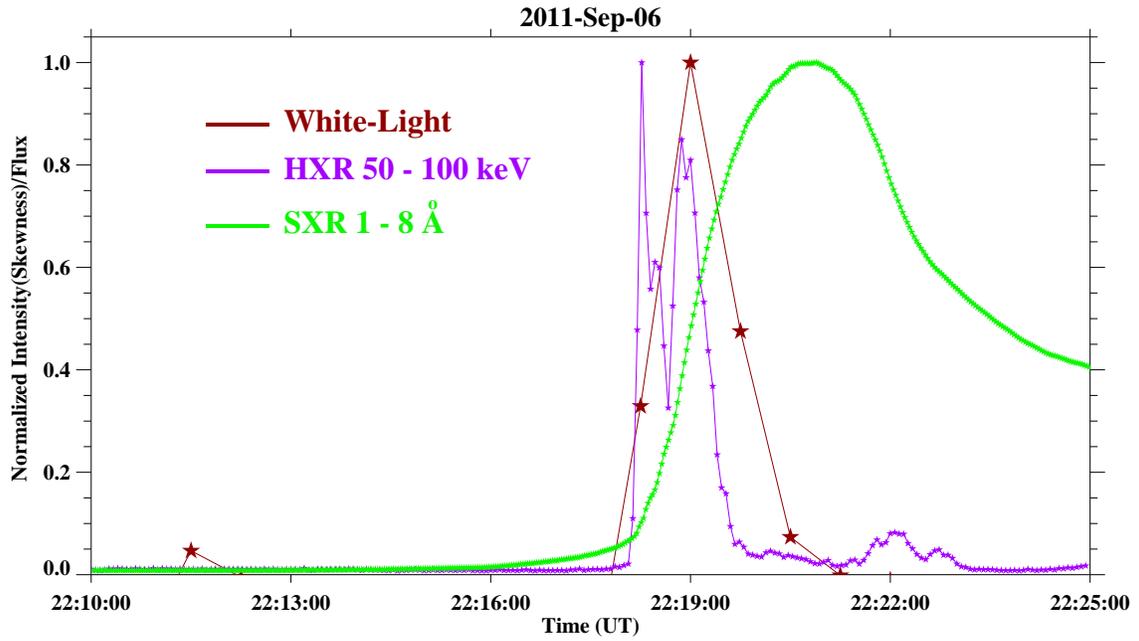}
\caption{Light curves of Flare I on 2011 September 6.
         Green curve: GOES SXR light curve in 1 - 8~\AA.
         Purple curve: RHESSI HXR light curve in energy band of 50 - 100 keV.
         Red curve with asterisks: HMI WL light curve.
         The cadence of HXR light curves is four seconds. The cadence of WL light curve is 45 seconds. All of the light curves are normalized to their peak counts. \label{lc}}
\end{figure}

\begin{figure}
\centering
\includegraphics[scale=0.88]{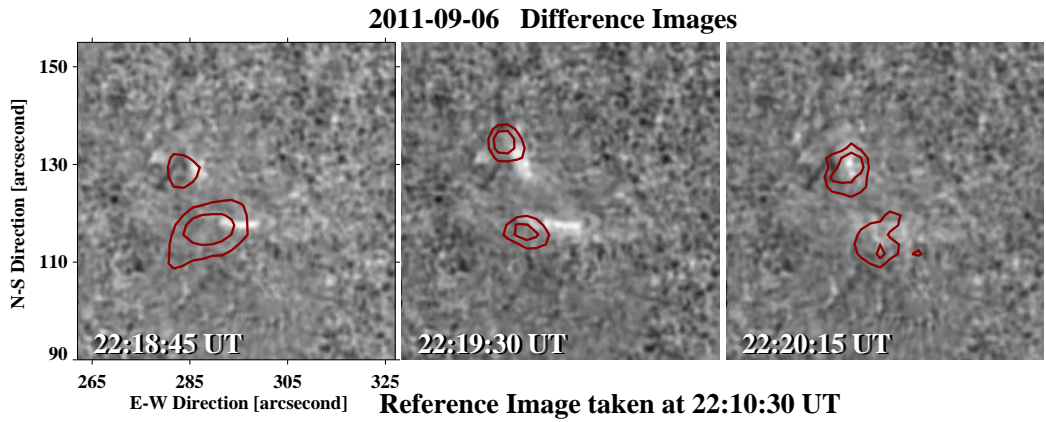}
\caption{Difference image in WL with HXR contours of Flare I. The WL images were taken at 22:18:45, 22:19:30 and 22:20:15 UT, respectively. On each WL image, the corresponding HXR contours (60\% and 80\%), in 50 - 100 keV, are plotted. The integration periods of HXR images are [22:18:30~UT $\sim$ 22:19:00~UT], [22:19:15~UT $\sim$ 22:19:45~UT] and [22:20:00~UT $\sim$ 22:20:30~UT]. This figure illustrates the spatial and temporal correlation between WL and HXR flare emission.   \label{WLHXR}}
\end{figure}

Figure~\ref{f1} shows HMI continuum images with a FOV (75\arcsec~ by 75\arcsec) covering the sunspot group in AR 11283. The upper-right panel was taken during the peak of the flare at 22:18:37 UT, on which the flare signal is not obvious in the raw data. By subtracting the reference image obtained before the flare (upper-left panel), we see two flare kernels on the difference image (lower-left panel). Unlike the UV or H$\alpha$ observations, the duration of continuum emission is much shorter. For this flare, the WL flare kernels can only be identified from five frames using the subtraction method. The light curve in WL is plotted together with the RHESSI HXR and GOES SXR light curves in Figure~\ref{lc}. In HXR, there are two major impulsive peaks three minutes apart. In the WL, creating a light curve is more complicated. Usually, the time sequence of difference images is obtained by subtracting a reference image. At this point, the selection of reference images is crucial. For instance, a light curve with a pre-flare reference could differ significantly from a light curve with a post-flare reference due to some non-flare variations. In addition, selection of ``average light curve'' or ``maximum light curve'' is arbitrary, the former represents the variation of overall emission and the latter represents the time profile of core emission \citep{Xu2006}. To avoid any randomness of selecting reference and reduce the uncertainties involved in the alignments and normalization, we adopt a method of generating light curves by using the high order moments \citep{Veronig2000}. In Figure~\ref{lc}, the third order moment (skewness), is used as a proxy of the intensity variation. We see one peak in the WL clearly but no fine structures due to the relatively low cadence (45 seconds) comparing to HXR's four-second time resolution. Nevertheless, the temporal correlation between WL and HXR is confirmed. To verify the spatial correlation between continuum and HXR sources, RHESSI CLEAN images are reconstructed using collimators 1 - 7 in an energy range of 50 - 100 keV. The time interval of each CLEAN image is 30 seconds and overlaps with the HMI observing time. Figure~\ref{WLHXR} shows the continuum images with the HXR contours, from which we see that the source locations are almost identical. The slight off-set is probably due to the projection effect because the formation heights are different for HXR and WL emission\footnote{For instance, an electron with 100 keV can reach a layer with a column density of $2.5 \times 10^{21}$cm$^{-2}$ (estimated using Equation 9 in \citet{Brown2002}). However the WL emission originates in the photosphere, where the column density reaches $10^{23}$cm$^{-2}$ \citep{Vernazza1981}. According to the VAL-F model \citep{Vernazza1981}, the height difference is at least 500 km.}. We conclude that at HMI's resolution of $0\arcsec.5$ per pixel, the HXR and WL sources occurred co-spatially and simultaneously. This result is expected as most of the previous observations have shown such a correlation \citep[e.g.,][]{Rust1975, Hudson1992, Xu2004b, MartinezOliveros2011}.

The WL emission reached its maximum at 19:19 UT. The strongest radiation came from the south kernel with a contrast enhancement of 24\%, which is somehow lower than the previous observations by \citep{Xu2004b}, in which the core enhancements were 45\% in the green continuum at 5200~\AA. Note that \citet{Xu2004b} observed an X10 flare which was much more energetic than this X2.1 flare. Consequently, one would expect stronger electron flux penetrating down to the lower atmosphere and generating intensive emission in the continuum during that X10 flare.

\begin{figure}
\centering
\includegraphics[scale=0.88]{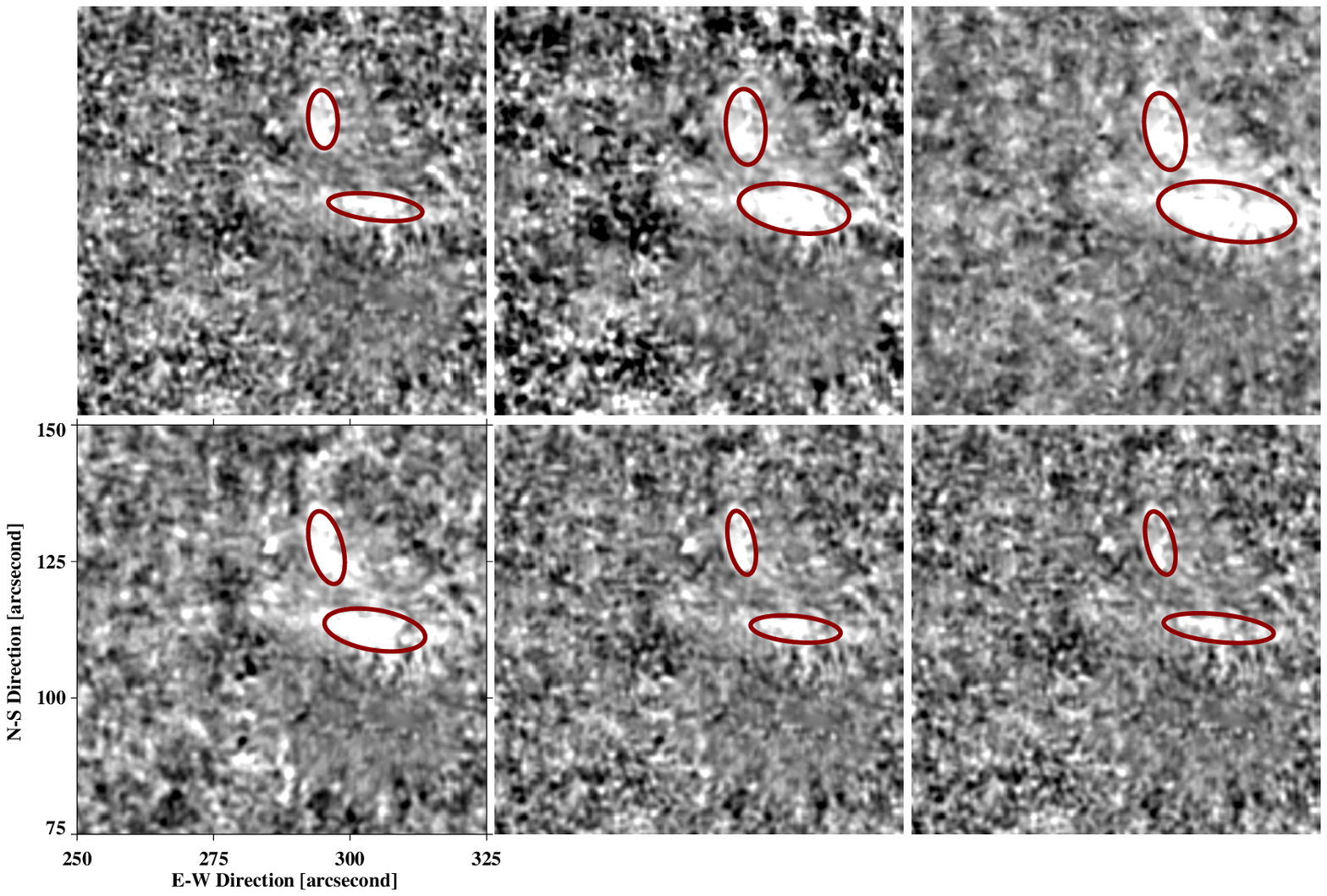}
\caption{Difference images of Flare I at six spectral positions, namely +174~m\AA, +103~m\AA, +34~m\AA, -34~m\AA, -103~m\AA, and -174~m\AA\ from the Fe I line center. The first row shows images in the red wing and the second row shows images in the blue wing. The contours represent the half maximum level from a two-dimensional Gaussian fitting. The size listed in Table~\ref{kernelsize1} is the FWHM of the minor axis.  \label{size}}
\end{figure}

During HXR and WL observations, we usually observe only one pair of the flare kernels, though exceptions are found in some special cases \citep{LiuC2007b}. This pair of conjugate footpoints could be the most intensive site of energy dissipation at a certain time. Figure~\ref{size} presents the flare kernels at the peak time in six spectral positions. The NRT spectral data is provided by the HMI team after proper alignment and normalization by their exposure time. Both flare kernels are fitted by a two-dimensional Gaussian function. Similar to \citet{Xu2012a}, the FWHM of the minor axis is calculated and defined as the `size' of each flare kernel. As we can see from Table~\ref{kernelsize1}, there is an obvious trend by which the source size increases toward the line center for both the north and south kernels. Such a wavelength-dependent size variation is consistent with the electron heating model as discussed by \citep{Xu2012a}.

\begin{table}[pht]
\caption{Characteristic size of flare kernels for Flare I. \label{kernelsize1}}

\begin{tabular}{lccccccr}
\\
\tableline\tableline

Spectral  & +172~m\AA & +103~m\AA & +34~m\AA & -34~m\AA & -103~m\AA & -172~m\AA  & Average \\
Position  &           &           &          &          &           &            &         \\
Size (N)  & 1.\arcsec14 & 1.\arcsec54 & 1.\arcsec55 & 1.\arcsec27 & 1.\arcsec01 & 1.\arcsec05  & 1.\arcsec26 $\pm$ 0\arcsec.24 \\
Size (S)  & 1.\arcsec00 & 1.\arcsec46 & 1.\arcsec87 & 1.\arcsec55 & 1.\arcsec14 & 1.\arcsec04  & 1.\arcsec34 $\pm$ 0\arcsec.34 \\
\tableline
\end{tabular}
\end{table}

\subsection{Flare II in comparison with Flare I}

\begin{figure}
\centering
\includegraphics[scale=0.88]{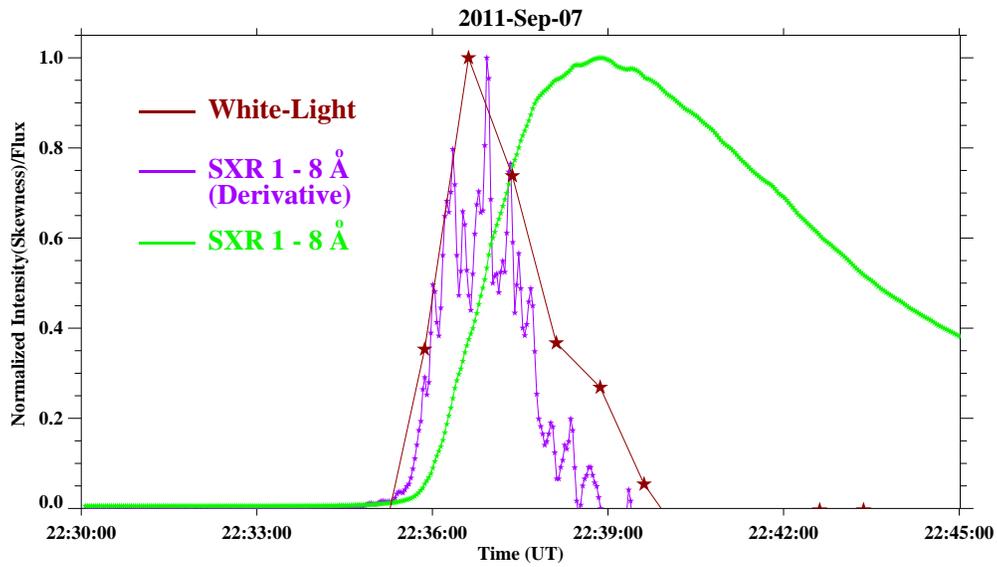}
\caption{Light curves of Flare II on 2011 September 7.
         Green curve: GOES SXR light curve in 1 - 8~\AA.
         Purple curve: Derivative of GOES SXR light curve in 1 - 8~\AA.
         Red curve with asterisks: HMI WL light curve.
         The cadence of WL light curve is 45 second. All of the light curves are normalized to their peak counts.\label{flare2lc}}
\end{figure}

\begin{figure}
\centering
\includegraphics[scale=0.88]{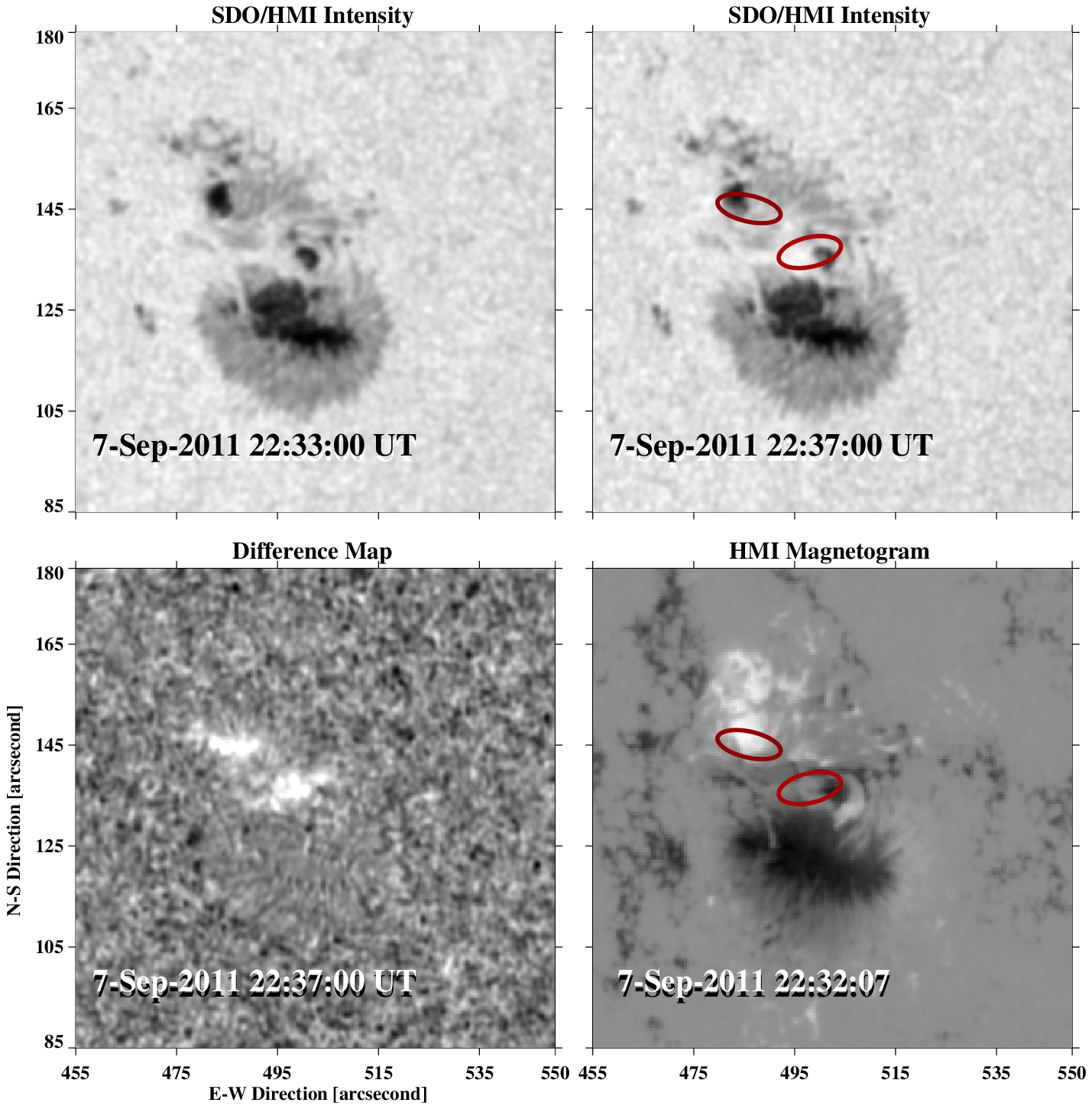}
\caption{WL images and Magnetogram of Flare II on 2011 September 7.
         Upper-left panel: HMI continuum image taken before the flare which is used as the reference frame.
         Upper-right panel: HMI continuum image taken during the flare.
         Lower-left panel: Difference image by subtracting the reference frame from the middle panel.
         Lower-right panel: HMI line-of-sight magnetogram taken before the flare at 22:32 UT. The red contours in two right panels show the positions of flare sources relative to the sunspot group.\label{flare2dif}}
\end{figure}

Figure~\ref{flare2lc} presents the light curve of Flare II, which is also a WL flare. The red curve with asterisks shows the temporal profile of HMI WL variation. Unfortunately, this flare occurred during RHESSI's night time and therefore HXR data is not available. We used the derivative of GOES SXR as a proxy for the HXR light curve. Again, similar to the flare I, we see that the WL emission is temporally correlated with electron precipitation. Figure~\ref{flare2dif} shows the WL images and flare signals using the subtraction method. The centroid of the flaring area at the peak time is around $N138\arcsec$, $W495\arcsec$. In the two right panels of Figure~\ref{f1} and Figure~\ref{flare2dif}, the 50\% contours of flare sources fitted using a two-dimensional Gaussian function, are plotted on the WL images. It is clear that the northern flare kernels of both flares are located directly above the same sunspot and the southern kernels reside at a `gap' area close to  the center of the sunspot group. Table~\ref{homologous} gives a comparison of the two homologous flares. Flare II was an X1.8 flare and relatively weaker than Flare I that has a GOES SXR class of X2.1. As a result, it is not surprising that Flare II has a relatively low contrast in the continuum.

\begin{table}[pht]
\caption{Comparison of the homologous flares. \label{homologous}}

\begin{tabular}{lcccccc}
\\
\tableline\tableline

Date & WL Intensity & WL           & WL          & X-ray     & Sunquake & Location \\
     & Enhancement  & Peak Time    & Duration    &      &          &          \\
     \tableline
Sep-06 & 24\%  & 22:19 UT     & $\sim$ 4 min     & RHESSI     & No       & N126\arcsec, W290\arcsec \\
            &       &              &             & GOES       &          &                          \\
Sep-07 & 20\%  & 22:37 UT     & $\sim$ 4 min     & GOES       & Yes      & N138\arcsec, W495\arcsec \\
            &       &              &             & derivative &          &                          \\
\tableline
\end{tabular}
\end{table}

\begin{figure}
\centering
\includegraphics[scale=0.88]{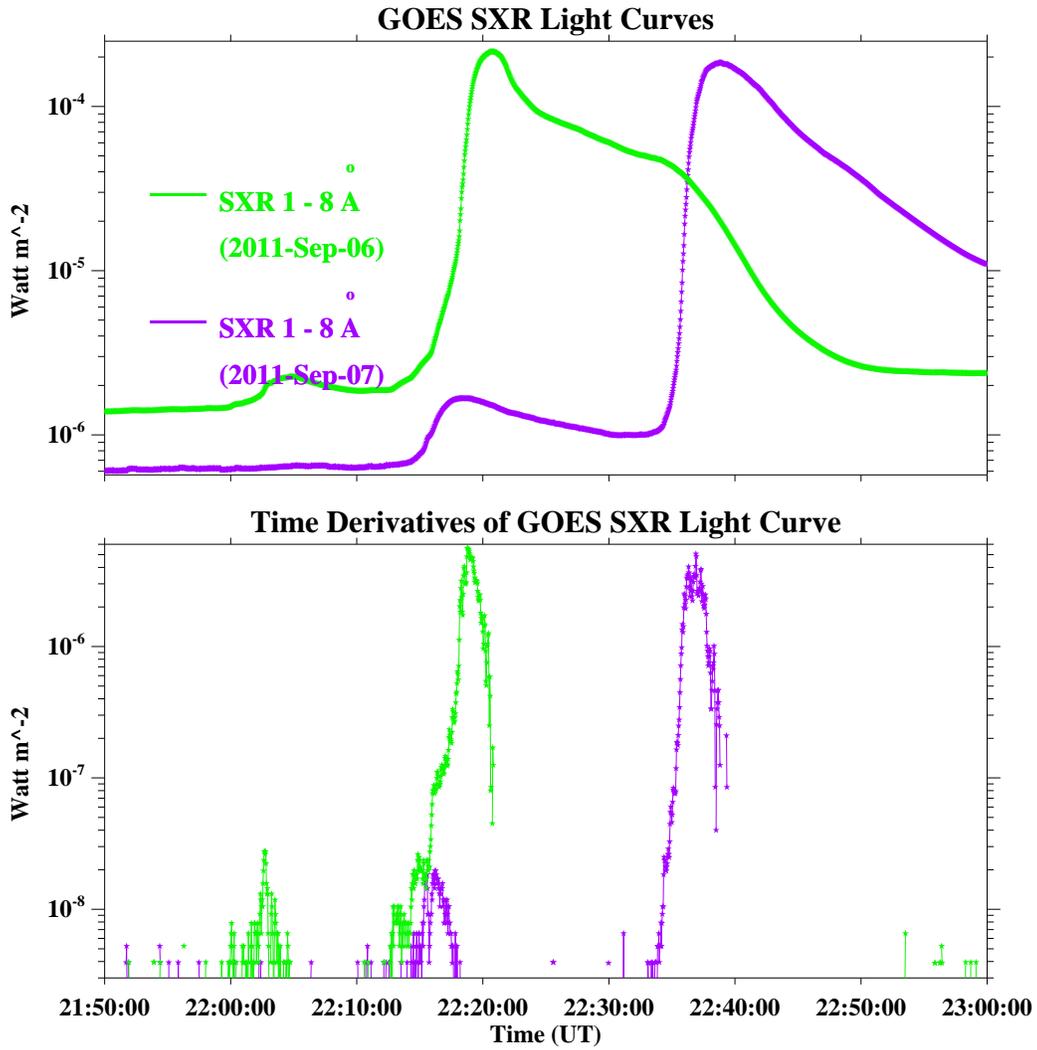}
\caption{Top panel: GOES SXR light curves for Flare I (green) and Flare II (purple).
         Bottom panel: Time derivatives of GOES SXR light curves for Flare I (green) and Flare II (purple).
         They have similar peak flux and time duration.  \label{goes2}}
\end{figure}

However, it is puzzling that the `weak' Flare II coincides with a clear sunquake \citep{Zharkov2013a}, which could not be identified during Flare I based on private communication with Dr. J. Zhao and Dr. S. Zharkov. Besides the seismic waves, there is no significant difference in emission between the two flares. In other words, the emission magnitudes and durations are similar in WL and HXR/SXR derivative (Figure~\ref{goes2}). The known possible causes of sunquakes include direct particle precipitation, backwarming, shock waves and the Lorentz force. The former three are associated with particle beams. Considering the structure of hosting AR and properties of accelerated particles, these two events should have similar seismic responses. Therefore, we suspect that besides the particle precipitation, there are some other combined effects in generating seismic emission, such as ambient atmospheric condition, and three-dimensional topology of surrounding magnetic fields, which are not well understood at present time.

\begin{figure}
\centering
\includegraphics[scale=0.88]{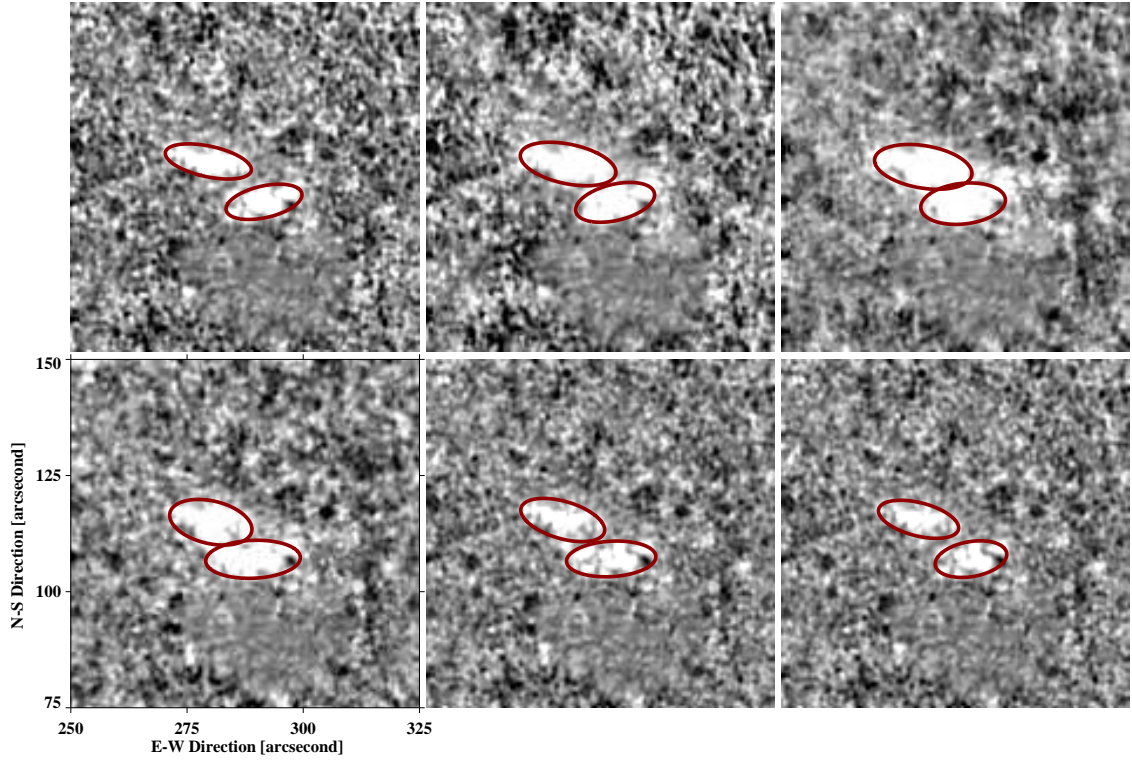}
\caption{Difference images of Flare II at six spectral positions, namely +174~m\AA, +103~m\AA, +34~m\AA, -34~m\AA, -103~m\AA, and -174~m\AA\ from the Fe I line center. The first row shows images in the red wing and the second row shows images in the blue wing. The contours represent the half maximum level from a two-dimensional Gaussian fitting. The size listed in Table~\ref{kernelsize2} is the FWHM of the minor axis.  \label{sizeII}}
\end{figure}

Figure~\ref{sizeII} presents the difference images of Flare II at six spectral positions as same as in Figure~\ref{size} for Flare I. Again, we see that the flare kernels are relatively compact in line wing and overspread in line center observations. Quantitative measurements of the characteristic sizes are listed in Table~\ref{kernelsize2}.

\begin{table}[pht]
\caption{Characteristic size of flare kernels for Flare II. \label{kernelsize2}}

\begin{tabular}{lccccccr}
\\
\tableline\tableline

Spectral  & +172~m\AA & +103~m\AA & +34~m\AA & -34~m\AA & -103~m\AA & -172~m\AA  & Average \\
Position  &           &           &          &          &           &            &         \\
Size (N)  & 1.\arcsec06 & 1.\arcsec37 & 1.\arcsec47 & 1.\arcsec48 & 1.\arcsec29 & 1.\arcsec15 & 1.\arcsec30 $\pm$ 0\arcsec.17 \\
Size (S)  & 1.\arcsec06 & 1.\arcsec22 & 1.\arcsec38 & 1.\arcsec29 & 1.\arcsec21 & 1.\arcsec20 & 1.\arcsec23 $\pm$ 0\arcsec.11 \\

\tableline
\end{tabular}
\end{table}

\section{Summary and Discussion}

In this paper, we studied two homologous X-class WL flares, which occurred on 2011 September 6 and 7 using HMI, RHESSI and GOES observations. We performed a detailed study of Flare I concerning several important aspects, and compared the emission properties of Flare I with Flare II. The findings are summarized and discussed as follows: \\

1. The continuum emission obtained by HMI was well correlated with RHESSI HXR observations in Flare I. Once again, this result confirms the close relationship between the WL emission and energetic electrons.

2. The maximum intensity enhancements were 24\% and 20\% for the twin flares, which are moderate comparing with previous WL observations, \citep[e.g.,][]{Lin1996, Xu2004b}. Note that the continuum around Fe I 6173~\AA\ was rarely used for flare studies prior to the launch of HMI/SDO, we have not established a comprehensive database for the WL flares and can not perform detailed statistical analysis until more flares are observed.

3. Using the 6-point data, the characteristic sizes of flare kernels were measured for both flares. At a certain time, for instance the flare maximum, the source size increased from the line wing to the line center. It is well known that the radiation from the line wing is formed lower than that from the line center. Therefore, the wavelength-dependent size variation indicates that the flare kernels are smaller at lower atmosphere than those at higher layers. This result favors the direct heating model as discussed in \citet{Xu2012a}.

4. Flares I and II have similar properties in WL but Flare I was not accompanied by a sunquake whereas Flare II was. In the literature, there are several models used to explain the physics of sunquakes: (1) Earlier theories \citep[e.g.,][]{Wolff1972} and observations \citep[e.g.,][]{Besliu-Ionescu2005, Donea2006, Zharkova2007} find close relationship between sunquake and energetic particles, which are also believed to responsible for WL flares. These energetic particles normally refer to electrons, as protons are much less in number and the $\gamma$-ray sources are found not co-spatial with WL and HXR flare sources \citep{Hurford2006}. The electron beams can penetrate down to photosphere and generate sunquakes directly as predicted by \citet{Wolff1972}. (2) On the other hand, these electrons may affect photosphere through a secondary effect, such as back-warming effect or shock waves, and produce sunquakes \citep{Donea2006, Hudson2008}. (3) Besides the electron-related models, there are some models that do not require electron beams to play an important role in producing sunquakes. \citet{Zharkov2011} analyzed the sunquake associated with an X2.2 flare on 2011 February 15. The authors found that the sunquake sources are located far away from the flare center. The discovery suggests that the erupting flux ropes may have an effect on photosphere and generates sunquakes. (4) Based on private communications with Dr. Donea, we learned that sometimes pre-flare heating can create a favorable environment, such as appropriate temperature and density, for sunquakes. With this hypothesis, a flare with pre-heating is more likely to be followed by a sunquake. (5) Another model that is not related to energetic particles is proposed by \citet{Hudson2008} assuming that the magnetic reconfiguration may lead to a sunquake.

In summary, we present two flares with similar pre-flare conditions and WL emission but only one flare is associated with the sunquake. We do not intend to distinguish a particular model from all the models discussed above. Instead, we conclude that the particle precipitation may not be the only cause of the sunquake associated with Flare II. There are some other effects that may also work together in generating seismic emission, such as an ambient atmospheric condition, and topology of surrounding magnetic fields, which are not well understood at present time.

\acknowledgements We thank the referee for valuable comments. Obtaining the excellent data
would not have been possible without the help of the HMI/SDO and RHESSI teams. YX thanks Dr. Sebastien Couvidat for providing the aligned HMI/SDO 6-point data. This work is supported by NSF-AGS-1153424, NSF-AGS-1250374, NSF-AGE-1153226, and NASA grants NNX13AG13G, NNX13AF76G and NNX11AQ55G.

\clearpage

\end{document}